\documentclass[12pt]{iopart}

\usepackage{graphicx}

\newcommand{\be}{\begin{equation}}
\newcommand{\ee}{\end{equation}}
\newcommand{\bea}{\begin{eqnarray}}
\newcommand{\eea}{\end{eqnarray}}

\newcommand{\nn}{\\ \nonumber}

\newcommand{\tl}{\tilde}

\begin{document}

\title{A simple approach to $\alpha$-decay fine structure}

\author{D.S. Delion$^{1,2,3}$, Monika Patial$^{4}$, R.J. Liotta$^{4}$, and R. Wyss$^{4}$}
\address{
$^1$"Horia Hulubei" National Institute of Physics and Nuclear 
Engineering, \\ 
407 Atomi\c stilor, POB MG-6, Bucharest-M\u agurele, RO-077125, Rom\^ania \\
$^2$Academy of Romanian Scientists,
54 Splaiul Independen\c tei, Bucharest, RO-050094, Rom\^ania \\
$^3$Department of Biophysics, Bioterra University, 81 G\^arlei str., Bucharest, RO-013724, Rom\^ania \\
$^4$Royal Institute of Technology (KTH), Alba Nova University Center, 
SE-10691 Stockholm, Sweden}

\begin{abstract}
We propose a simple method to evaluate  
$\alpha$-transition rates to low-lying excited states in even-even
nuclei. For this a realistic $\alpha$-daughter double folding interaction 
is approximated by a parabola in the region where the decay process takes
place. This allows us to evaluate the penetration probability analytically.
The main experimental features of branching ratios to excited states
are reproduced by this simple approach.
\end{abstract}



\maketitle

\section{Introduction} 
\label{sec:intro} 
\setcounter{equation}{0}
\renewcommand{\theequation}{1.\arabic{equation}} 

Relative $\alpha$-decay rates are often  explained in terms of a preformed 
$\alpha$-particle  which penetrates through the Coulomb barrier \cite{Gam28}.
However the calculation of absolute half lives requires a knowledge of the
formation probability of the $\alpha$-particle on the surface of the mother
nucleus. This is a very difficult task. Therefore the formation probability
is often replaced by effective parameters. Such parametrizations are rather
successful in describing decay widths and, therefore, they are use
extensively in $\alpha$-decay analyses (see e.g. Ref. \cite{effec} and references therein). 
However, there are recent high precision measurements of relative decay rates (fine structure)
from even-even emitters to excited states in the corresponding daughter nuclei
\cite{Ako98,Den09} which do not involve absolute decay widths.
Their description is usually given within a rather involved coupled channels method
by using a double-folding $\alpha$-core potential \cite{Del15}.  
This formalism is able to reproduce $\alpha$-decay intensities with a good accuracy.

The aim of this paper is to explain the gross features of the fine structure 
experimental data using a much simpler analytical approach.
In Section 2 we give the necessary theoretical details how to estimate
partial $\alpha$-decay widths, in Section 3 we systematize the experimental
material concerning the $\alpha$-decay fine structure from even-even emitters
and in the last Section we draw Conclusions.

\section{Theoretical background} 
\label{sec:theor}
\setcounter{equation}{0}
\renewcommand{\theequation}{2.\arabic{equation}} 

\begin{figure}[ht] 
\begin{center} 
\includegraphics[width=7cm,height=8cm]{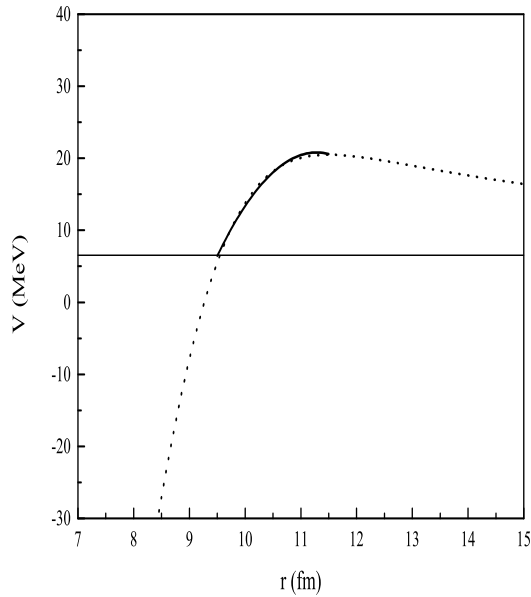} 
\caption{
Double folding potential (dashed line) and the 
parabolic fit of the internal region (solid line).
}
\label{fig1}
\end{center} 
\end{figure}

We start by noticing that a realistic 
$\alpha$-nucleus spherical potential reproducing well
scattering data is given by the double folding procedure using the M3Y 
nucleon-nucleon interaction  \cite{Ber77,Sat79,Car92}. 
The most relevant part of this potential in the case of 
$\alpha$-decay is the
region  between the innerst and outermost turning points, We found that in
this region the potential can very well be approximated by the expression,
\bea
\label{pot}
V(R)&=&c-a(R-R_0)^2~,~~~R\leq R_m
\nn
&=&\frac{2Ze^2}{R}~,~~~R> R_m~,
\eea
where Z is the charge number of the daughter nucleus. The parabolic form
of the potential in the first line is defined by the parameters $R_{0}$, $c$ and $a$.
These parameters are found by a fitting procedure
and the matching radius $R_m$ is determined by imposing the
continuity of the potential. 
An example of the quality of the fitting procedure can be seen in Fig. \ref{fig1}.

We will apply our method to $\alpha$-decay from even-even emitters to yrast
states in the daughter nuclei for which there are
experimental data \cite{Ako98}. We found that in these cases the parameters
of the analytic potential (\ref{pot}) are approximately given by,
\bea
\label{par}
R_0&=&3.24A^{0.234}~,~~~\sigma=0.0009
\nn
c&=&0.97\frac{2Ze^2}{R_0}~,~~~\sigma=0.0054
\nn
a&=&1.76A^{0.177}~,~~~\sigma=0.0198~,
\eea
where $Z$ ($A$) is te charge (mass) number of the 
daughter nucleus. The standard errors $\sigma$ are also given.

We will evaluate the partial decay widths using the spherical semiclassical
approximation, which is known to be 1-2\% accurate  with
respect to the exact solution \cite{Del10}.
The action integral for the nuclear interaction using the expression
Eq. (\ref{pot}) can now be evaluated analytically. 
We found out that $R_m\approx R_0+0.3$ fm. We will evaluate the action integrals
in the inner and out intervals divided by the radius $R_0$.
Thus, the inner part becomes 
\bea
\label{Kint}
&&K_{int}(Q)=\int_{R_1}^{R_0}\sqrt{\frac{2\mu}{\hbar^2}[-a(R-R_0)^2+c-Q]}dR
\nn
&=&\frac{1}{2}\sqrt{\frac{2\mu}{\hbar^2}}(R_0-R_{1})\sqrt{-a(R_0-R_{1})^2+c-Q}
\nn
&+&\sqrt{\frac{2\mu}{\hbar^2}}\frac{(c-Q)}{2\sqrt{a}}
tan^{-1}\left(\frac{\sqrt{a}(R_0-R_{1})}{\sqrt{-a(R_0-R_{1})^2+c-Q}}\right)~,
\eea
where $Q$ is the Q-value of the process and $R_1$ is the innerst turning point,
which has the value
\bea
\label{R1}
R_1=R_0-\sqrt{\frac{c-Q}{a}}~.
\eea
A similar action integral for the outer part gives
\bea
K_{ext}(\chi,\rho_0)&=&
\int_{\rho_0}^{\chi}\sqrt{\frac{\chi}{\rho}-1}d\rho
\nn&=&\chi\left(cos^{-1}
\sqrt{\frac{\rho_0}{\chi}}-\sqrt{\frac{\rho_0}{\chi}-1} 
\sqrt{\frac{\rho_0}{\chi}}\right)~,
\eea
where the Coulomb parameter $\chi$ and reduced radius $\rho$ are given by,
\bea
\chi&=&\frac{4Ze^2\mu}{\hbar\kappa}
\nn
\rho&=&\kappa R~,~~~\kappa=\frac{\sqrt{2\mu Q}}{\hbar}~.
\eea
The partial decay width to the excited state with spin $J$ and 
excitation energy $E_J$
is proportional to the exponent of the sum of the two action integrals \cite{Del10},
\bea
\label{GamJ}
\Gamma_J^{(th)}&=&\exp\left\{-2\left[K_{int}\left(Q-E_J-\frac{\hbar^2J(J+1)}{2\mu\tl{R}^2}\right)
\right.\right.\nn&+&
\left.\left.K_{ext}(\chi_J,\kappa_JR_0)
+\frac{J(J+1)}{\chi_J}\sqrt{\frac{\kappa_JR_0}{\chi_J}-1}\right]\right\}~,
\eea
where the centrifugal potential was evaluated at 
\bea
\tl{R}&=&\frac{R_1+R_0}{2}=R_0-\frac{1}{2}\sqrt{\frac{c}{a}}~.
\eea
and the channel values of the Coulomb parameter and momentum, respectively,
are given by,
\bea
\label{chan}
\chi_J&=&\frac{4Ze^2\mu}{\hbar\kappa_J}
\nn
\kappa_J&=&\frac{\sqrt{2\mu(Q-E_J)}}{\hbar}~.
\eea
The total decay width is given by the sum of the corresponding partial widths,
i. e.,
\bea
\label{Gam}
\Gamma^{(th)} &=& \sum \Gamma_J^{(th)}~.
\eea
Let us stress on the fact that the parameters of the interaction potential (\ref{par})
describe scattering data of $\alpha$-particles. 
Thus, the $\alpha$-particle is supposed to exist with the unity probability. 
In order to estimate total and partial $\alpha$-decay formation probabilities 
we define total and partial spectroscopic factors respectively, as, 
\bea
\label{S}
S&=&\frac{\Gamma^{(exp)}}{\Gamma^{(th)}}~,~~~
S_J=\frac{\Gamma_J^{(exp)}}{\Gamma_J^{(th)}}~.
\eea

\section{Numerical application} 
\label{sec:appl} 
\setcounter{equation}{0}
\renewcommand{\theequation}{3.\arabic{equation}} 

\begin{figure}[ht] 
\begin{center} 
\includegraphics[width=7cm,height=8cm]{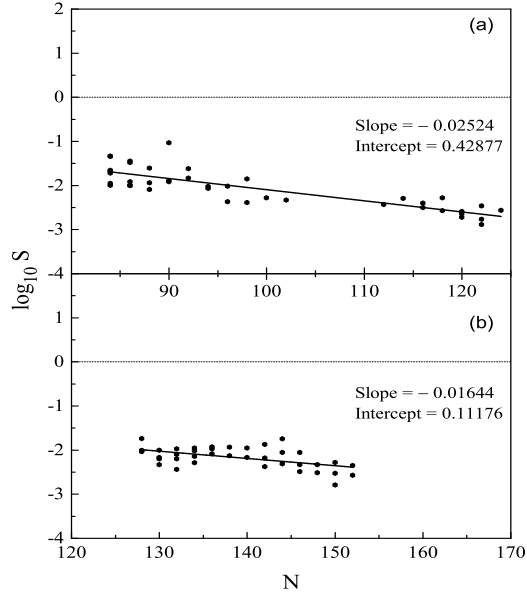} 
\caption{
(a) Logarithm of the spectroscopic factor versus the neutron number for $N<126$.
By a solid curve we plotted the fitting line.
(b) Same as in (a) but for $N>126$.
}
\label{fig2}
\end{center} 
\end{figure}

We analyzed available experimental data concerning $\alpha$-decays to excited states
in even-even nuclei \cite{Del15}.
In Fig. \ref{fig2} we plotted the total spectroscopic factor as a function
of the neutron number
for (a) $N<126$ and (b) $N>126$. One can see in this Figure the striking
feature that the logarithm of the spectrocpic factor follows two separate
lines  depending upon whether the neutron number is larger or smaller than the
magic number $N=126$.

\begin{figure}[ht] 
\begin{center} 
\includegraphics[width=7cm,height=8cm]{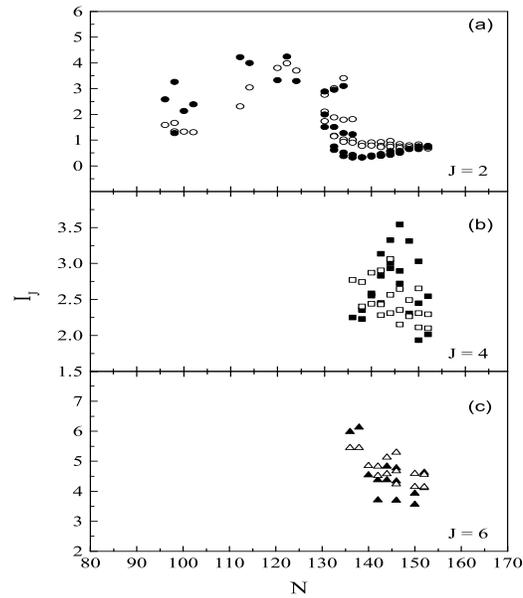} 
\caption{
Intensity (\ref{IJ}) versus neutron number for $J=2$ (a) $J=4$ (b) and $J=6$ (c).
}
\label{fig3}
\end{center} 
\end{figure}

\begin{figure}[ht] 
\begin{center} 
\includegraphics[width=7cm,height=8cm]{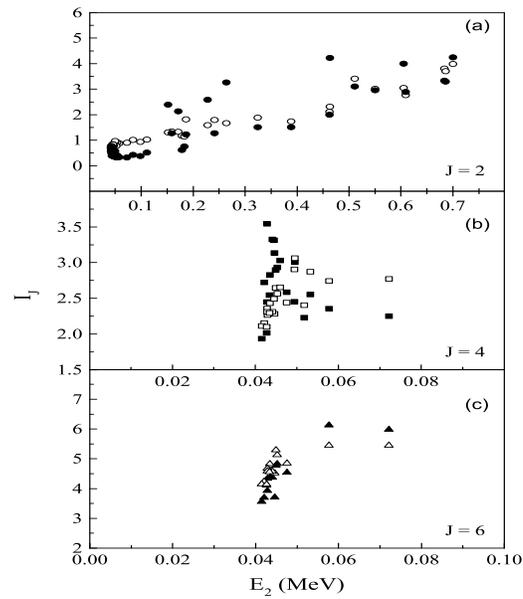} 
\caption{
Same as in Fig. \ref{fig3}, but versus the excitation energy $E_2$.
}
\label{fig4}
\end{center} 
\end{figure}

It is convenient to analyze the $\alpha$-decay fine structure in terms of
the so-called decay intensities \cite{Del06}
\bea
\label{IJ}
I_J=\log_{10}\frac{\Gamma_0}{\Gamma_J}~.
\eea
In Fig. \ref{fig3} we plotted the intensities defined by the above relation
versus the neutron number corresponding to  (a) $J=2$, (b) $J=4$ and (c) $J=6$. 
In this Figure the experimental values are represented by dark symbols while 
open symbols correspond to the results of our calculations.
One sees that the experimental features are reasonable well reproduced by 
the theoretical estimates. In Fig. \ref{fig4} we plotted the same values, but 
as a function of the excitation energy $E_2$. One notices
the linear increasing trend of the intensity $I_2$, as predicted in 
Ref. \cite{Del09}.

In order to avoid the exponential influence of the penetrability in the
decay process one usually defines
the hindrance factor (HF) as \cite{wau},
\bea
\label{HFJ}
HF_J=\frac{S_0}{S_J}=\frac{\Gamma_0^{(exp)}}{\Gamma_J^{(exp)}}\frac{\Gamma_J^{(th)}}{\Gamma_0^{(th)}}~.
\eea
The logarithm of this quantity can be written as a difference between 
experimental and theoretical decay intensities, i. e.
\bea
\label{logHFJ}
\log_{10}HF_J=I^{(exp)}_J-I^{(th)}_J~.
\eea
This difference characterizes the other elements which we neglected
in our simple approach, namely the deformation, given by the
coupling between channels, and the clustering probability in the decay process. 

\begin{figure}[ht] 
\begin{center} 
\includegraphics[width=7cm,height=8cm]{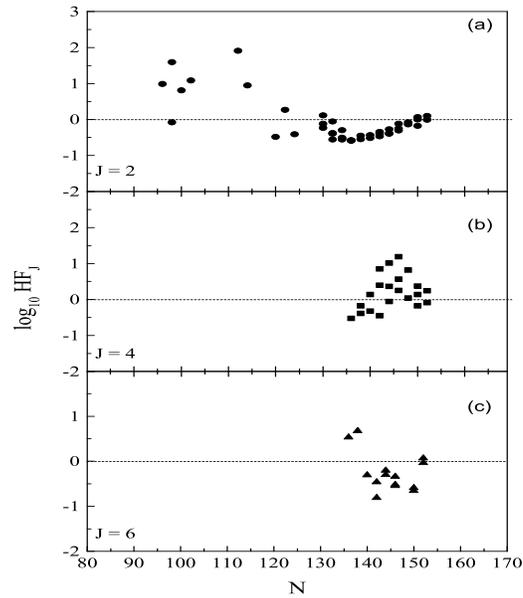} 
\caption{
Logarithm of the hindrance factor (\ref{HFJ}) versus neutron number 
for $J=2$ (a) $J=4$ (b) and $J=6$ (c).
}
\label{fig5}
\end{center} 
\end{figure}

In Fig. \ref{fig5} we plotted $\log_{10}HF_J$ as
a function of the neutron number. For the states $J=2^+$ one sees that 
the fine structure can be reproduced rather well 
in the region $N>126$, while for $N<126$ the the above mentioned
features are necessary in order to explain experimental data.
For the transitions to the states $J=4^+$ one sees a maximum in the region of
Pu isotopes, as already found in Ref. \cite{Del06}. This is connected to
a subshell effect \cite{Buc12}.

\begin{figure}[ht] 
\begin{center} 
\includegraphics[width=7cm,height=8cm]{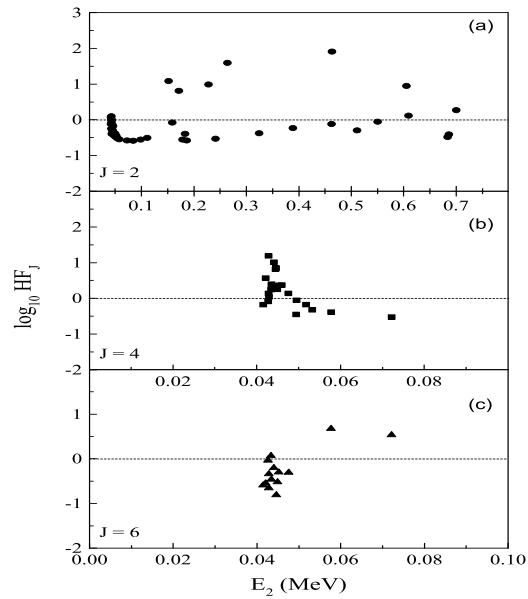} 
\caption{
Same as in Fig. \ref{fig5}, but versus the excitation energy $E_2$.
}
\label{fig6}
\end{center} \end{figure}

In Fig. \ref{fig6}, the hindrance factor is also shown as a function of 
the excitation energy $E_2$. As mentioned, the theoretical deviations are
to be regarded as due to the influence of deformation and $\alpha$-clustering
upon the fine structure. 
 
\section{Conclusions} 
\label{sec:concl} 
\setcounter{equation}{0}
\renewcommand{\theequation}{4.\arabic{equation}} 

In conclusion, we have analyzed the $\alpha$-decay fine structure to low-lying 
excited states in even-even nuclei by using an analytical semiclassical approach.
We approximated the realistic $\alpha$-daughter double folding interaction
by a parabola in the spatial region which is relevant in the decay process.
Partial decay widths were estimated by using standard spherical semiclassical
approach where the action integrals have close analytical forms.
By analysing hindrance factors, we found that the main experimental features are
reproduced by this simple method within one order of magnitude.
Further improvement is due to the deformation effects induced
by the coupling between multipoles and $\alpha$-particle formation probabilities.

\vskip5mm
{\bf Acknowledgments}
\vskip5mm

This work was supported by the Royal Institute of Technology, Stockholm,
the Swedish Research Council (VR) under Grants No. 621-2012-3805 and No. 621-2013-4323,
and by the Grants of the Romanian National Authority for Scientific Research, 
CNCS-UEFISCDI, PN-II-ID-PCE-2011-3-0092 and PN-09370102.

\end{document}